# HVACKer: Bridging the Air-Gap by Attacking the Air Conditioning System

Yisroel Mirsky, Mordechai Guri, and Yuval Elovici

Ben-Gurion University of the Negev, Beersheba, Israel

{yisroel, gurim, elovici} @ post.bgu.ac.il

*Abstract*—**Modern corporations physically separate their sensitive computational infrastructure from public or other accessible networks in order to prevent cyber-attacks. However, attackers still manage to infect these networks, either by means of an insider or by infiltrating the supply chain. Therefore, an attacker's main challenge is to determine a way to command and control the compromised hosts that are isolated from an accessible network (e.g., the Internet).**

**In this paper, we propose a new adversarial model that shows how an air gapped network can receive communications over a covert thermal channel. Concretely, we show how attackers may use a compromised air-conditioning system (connected to the internet) to send commands to infected hosts within an air-gapped network. Since thermal communication protocols are a rather unexplored domain, we propose a novel line-encoding and protocol suitable for this type of channel. Moreover, we provide experimental results to demonstrate the covert channel's feasibility, and to calculate the channel's bandwidth. Lastly, we offer a forensic analysis and propose various ways this channel can be detected and prevented.**

**We believe that this study details a previously unseen vector of attack that security experts should be aware of.**

## I. INTRODUCTION

THERE is a common defense strategy in which an air gap (some physical gap) is placed between networks that maintain sensitive systems and all other nonessential infrastructures which are connected to a public network. The thinking behind this strategy is: if there are no connections between the sensitive network and all other public networks (e.g. the internet) then the sensitive network is secure from remote attacks. In other words, the belief is that attackers cannot breach an air gapped network if they have no physical way of communicating with it. However, this strategy is not foolproof. As a result, attackers have been developing methods for "bridging the air gap," and in several cases they have been successful [1–3].

A well-known attack in which an air gap was bridged occurred when the computer virus "Stuxnet" targeted the Iranian nuclear program [4]. This virus was able to intrude an isolated computer network that operated delicate centrifuges. Another example is the "agent.btz" [5], a malware which was used against US military networks. According to publications it caused no damage, however, it was the first confirmed case of the infection of a secure US government network which was separated by an air gap.

### A. Advanced Persistent Threats

In many cases, it is highly beneficial for an attacker to be able to send commands to malware it has successfully planted in an air gapped network. In order to understand why, it is important to understand what advanced attack looks like. Advanced attacks, also known as Advanced Persistent Threats (APTs) [6–8], are attacks that involve a process of several stages before arriving at their ultimate goal. These attacks are "Advanced" in terms of their technological awareness, "Persistent" in their well-supported campaign to achieve the final goal, and considered to be "Threats" to the intended victim as the attackers have both capability and intent. The goal of the APT depends on what the attacker desires to achieve. Many APTs, such as those that occurred to RSA [9] and various diplomatic agencies [10], have the goal of data exfiltration (data theft), while some, such as "Stuxnet", have the goal of causing damage to assets within the target network.

In Fig. 1, the stages of an APT are illustrated. In the first stage, the attacker researches the target network in order to understand how to gain access to it, and to perform the attack. During this stage the attacker typically researches the network's equipment, as well as researches the target organization's employees. Knowledge on the employees is important in order to perform stage 2, since the initial intrusion into the network is typically accomplished via a social networking attack. An example of a social engineering attack is where the attacker sends an email with an infected attachment, and the attachment is opened by the unsuspecting victim, thereby stealthily infecting the victim's computer with the attacker's malware. Another example of a social engineering attack is where the attacker creates an infected USB thumb drive that loads the attacker's malware into any computer to which the drive is plugged into. The attacker then gets the unsuspecting victim to plug the drive into their computer by curiously placing the drive in various locations near the victim (such as on the floor).

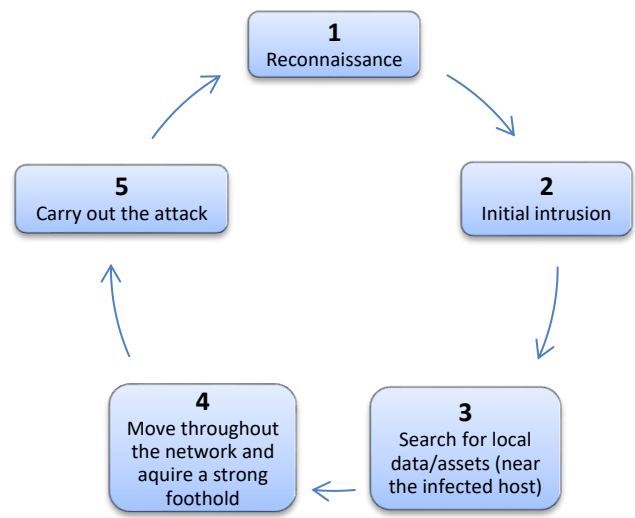

Fig. 1. The stages of an APT on a victim's network. Typically, stages 2-5 are controlled by the attacker from a remote location over a public network such as the internet.

In stage 2, at least one computer in the target network has been infected with the attacker's malware. At which point stage 3 begins where the malware searches the infected host (computer in a network) for ways to gain access to higher privileged resources and data (that can be access through various exploitations in the host's system's code). Once sufficient control has been acquired over the host, stage 3 begins where the malware begins spreading across the network to strengthen its foothold on the network, and acquire access to the attacker's target asset within the network (in most cases the initial intrusion is not topologically near the attack's target asset. A target asset can be a server, subsystem, database… etc. depending on the goal of the attack.

Once the asset has been reached, or at a determined time, the attack is performed and typically all evidence of the APT is self-removed by the infected hosts from the network.

The stages in Fig. 1 are placed in a cycle, since in some cases the attacker can perform multiple sequential attacks, each different than the last. In which case, the attacker may need to perform stages 1-4 again, although these stages will be different that the last attack (perhaps shorter of non-existent).

The steps of an APT attack become challenging for the attacker when the targeted network is isolated by an air gap. After stage 2, the attacker can no longer control the activities of the malware that now runs inside the target network. This generally results in an autonomous attack like in the case of Stuxnet. In order to be more effective, the attacker would benefit from the ability to command and steer the APT attack throughout the process. For example, in the case where the attacker's goal is data exfiltration, the attacker may change the target file types or content that should be found and collected for exfiltration. In the case where the attacker's goal is a cyber-attack, (i.e. goal is to destroying/modify some asset(s) in the compromised network) the attacker may want to control the exact time when the destruction/modification process should begin.

To conclude, an APT involves an advanced attacker who is extremely motivated, and would benefit from a command and control channel to the compromised hosts within the target network [7].

### B. Exploitation of Local Networks

In some cases, an insecure / publically connected network overlaps the same physical space as an air gapped network. These networks can easily be overlooked by security experts and then later compromised as a stage in attacking the nearby secure network.

One example of such an insecure and seemingly innocent network and asset is the heating, ventilation, and air conditioning (HVAC) system that is connected to an accessible network or even to the Internet. Today, many large buildings' HVAC systems are connected together by a network in order to report failures and to provide controls of their activity (such as setting the temperature in every room) [11].

With the boom of the Internet-of-Things, these types of infrastructures are being connected to publically accessible networks [12].

While there are different vendor specific TCP/IP connected centralized air conditioning systems [13], a specific example of a network-based building management system which handles HVAC is BACnet (building automation and control networks) [14]. BACnet is an ANSI and ISO standard communications protocol used to communicate with control systems for applications such as heating, lighting and fire detection. A BACnet network can be connected to the internet for remote management or as a means for connecting several BACnet systems together [11], [15]. A BACnet internet gateway can be compromised to give an attacker control over a building's HVAC system.

An example of a similar system being compromised occurred in 2013 when two researchers hacked into a Google office building's HVAC system with little difficulty [16]. The compromised system (Tridium Niagara AX platform) not only gave the researchers access to the heating and cooling of targeted rooms, but it also gave them access to floor plans and possibly bridged LAN hosts (computers in the local network). Similar vulnerabilities have been seen in this system, giving attackers access to building locks, electricity, elevators, and various other automated systems of the building [17]. At the time of writing this paper (2015), there are currently 36,287 Niagara web portals that can be found using the Shodan computer search engine [18], 27,657 of which are in the U.S. alone. This is an increase of 15,743 exposed systems since 2012, and only 269 of the systems are protected with HTTPS [19].

Lastly, another possibility for compromising an HVAC system (specifically the air conditioning section) is to attack a Wi-Fi-enabled thermostat [20]. Many of these types of interfaces are susceptible to Wi-Fi or other social engineering attacks [21].

### C. Bridging and Air Gap using Thermal Signals

In this study, we propose a new adversarial model that enables a remote attacker to send commands to infected hosts over an air gap: The attacker compromises an HVAC network located on the same premises as the infected hosts, and then uses the air conditioning system to send thermal signals to them. The thermal signals are picked up by the infected hosts' internal thermal sensors. This model essentially establishes a one directional covert broadcast channel to infected hosts within an air gapped network. Such a broadcast system could be used to manage APT remotely, or for some other attack such as initiating a denial of service (DoS) attack performed in order to interfere with daily work [22].

Studies on the formation of covert channels based on temperature effects have been conducted before. For instance, in [23] and [24] the authors show how a covert channel can be created based on clock skewage from imposed CPU loads. However, this type of channel is not possible over an air gap due to the lack of a wired interconnectivity. Moreover, in [25] the authors mention how two air gapped servers in a stack could possibly transfer bits to one another by heating up the environment around them. However, no details or concrete experimental results were provided.

A recent work of ours [26] we showed how two computers can communicate with each other using thermal signals by heating up their environment by imposing heavy computational loads. In contrast, our work provides a channel capable of significantly higher bit rates (40 vs 7.5 bits/sec), and proposes a new vector of attack that does not require the incidental placement of two PCs next to each other (one on an air gapped network, the other not).

Therefore, the contributions of this paper are as follows:

1) **A new adversarial attack model for bridging an air gap**. We propose a new adversarial attack model which can be used to send commands to infected hosts within an air gapped network. We discuss the steps required by the attacker to establish this attack, as well as the forensic evidence which results from using this model. We also propose methods for detecting this attack model, and steps that can be taken to prevent it. Moreover, we provide a high level protocol that could be used to manage the proposed attack. Awareness of this example protocol may assist security experts in revealing covert thermal channels.

2) **A thermal environment line-encoding**. We propose a line-encoding (baseband modulation) which can be used to create a narrow-band communication channel over an open air environment. The proposed channel is used to measure the feasibility of the attack model as a signaling channel. We provide experimental results to demonstrate the channel's usage in the attack model, and discuss the secrecy of the proposed protocol and line-encoding.

The rest of the paper is structured as follows: In Section 2, a motivating attack scenario is proposed. In Section 3, the channel with its theory and line-encoding is detailed. In Section 4, a thermal transmission protocol is proposed, along with a more advanced management protocol. In Section 5, experimental results are presented. In Section 6, Forensics and Countermeasures are discussed, and in Section 7, a conclusion with prospects for future work is provided.

## II. Attack Scenario

In this section we detail plausible attack scenario, in which the proposed adversarial model may be used to an attacker's advantage.

### A. The Motivating Scenario

Let us assume that the target organization uses a contained Ethernet environment, physically disconnected from the Internet and every other public network. However, in parallel to the organization's network, the encompassing building has a heating, ventilation, and air conditioning (HVAC) system whose control center has a connection to some other public network such as the Internet (see Fig. 2).

In this scenario, the attacker wishes to perform an APT on this organization, and plants malware on random hosts within the network. The attacker infects some of hosts behind the air gap using one of numerous social engineering attacks. For instance, by means of an infected USB device plugged in by an insider [27]. USB infection is a common way to make an initial infection to an air-gapped network and to spread malware to new hosts. This is the technique that was used by Stuxnet, Flame and Gauss [28]. Furthermore, attacks on the supply-chain can be used to implant the malware on the hosts like the case of the NSA secretly planting backdoors on routers shipped abroad [29]. It is reasonable to assume that other air-gapped networks have been bridged with similar tactics, although they have not been publically reported.

The attacker in this scenario needs the flexibility of steering the attack by activating the actions in Table I when needed. For instance, the attacker may want to search for and edit a file in the network several years later (other examples can be found in Table I). In order to accomplish this, the attacker needs the ability to covertly broadcast commands to the infected hosts behind the air gap. Therefore, the attacker plans this ahead accordingly, and uses the adversarial attack model propose in this paper.

### B. The Proposed Adversarial Attack Model

The overview of the proposed adversarial attack model can be seen in the use-case (Fig. 2) and is performed as follows. Step 1: The attacker's commands are sent to a command & control server somewhere in the internet. This is done to add a degree of separation between the attacker and the attack thereby protecting the attacker's identity [7]. Steps 2-3: The server forwards the commands to the infected HVAC system's central management unit via the system's internet web interface (there for legitimate remote control over the HVAC system [13], [17]). Steps 4-5: The HVAC system transmits the commands over the air by altering the thermal environment according to a predetermined protocol.

This attack model conforms to an APT attack in the following way. During the reconnaissance stage of the APT (stage 1 in Fig. 1), the attacker determines what type of HVAC system or building management protocol the organization is

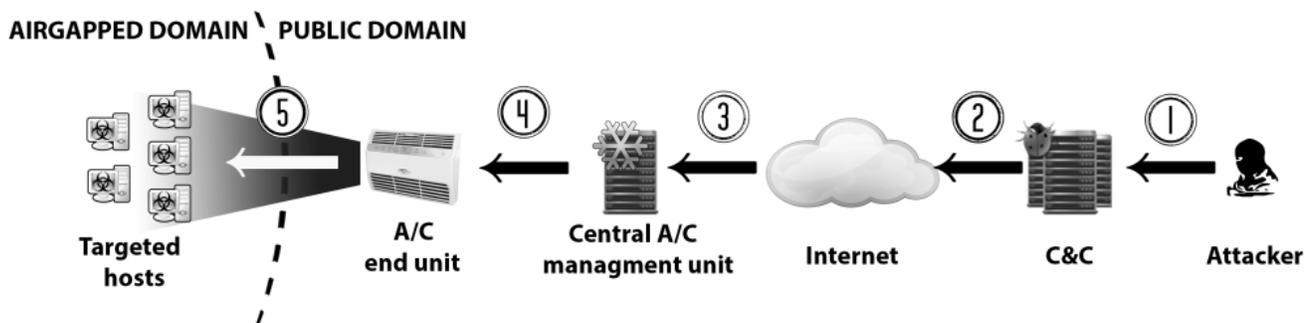

**AIRGAPPED DOMAIN  PUBLIC DOMAIN**

Targeted hosts — A/C end unit ⑤ — Central A/C managment unit ④ — Internet ③ — C&C ② — Attacker ①

Fig. 2. The use-case of the proposed adversarial attack model. The compromised hosts within an isolated network receive commands from the attacker over the air gap (literally and figuratively).

TABLE I
EXAMPLES OF OPPORTUNISTIC ATTACKS THAT MAY BE PERFORMED ON A
TARGET NETWORK

| Task | Description |
|---|---|
| *Search and delete a file* | The removal of sensitive files for the advantage of some cause. For instance, military intelligence documents or digital evidence that support a certain case. This can be done by searching for keywords. |
| *Search and edit a file* | The changing of files or data entries found by a key-word similarity search. For instance, replacing names of people or locations or other string swaps that violate a file's integrity. |
| *Temporarily disable a system asset* | Disabling a host, server or subsystem by means of an internal DoS attack or the direct intervention of some infected host. |
| *Temporarily disable a security protocol* | Disabling the security measures of a host, subsystem or even a security authentication waypoint for personnel. |
| *Move collected data to a staging area* | In the case where sensitive information has been gathered by the infected hosts, the data can be copied to a common extraction point. For instance, an insider whose presence within the isolated network is only temporary. |
| *Self-Destruct* | Covering tracks. All evidence of the infection and log files that may indicate the existence of the APT are deleted. |
| *Encryption key change* | The changing of the encryption key is used in the communication between infected hosts or their encrypted log files. |

using and maps out its network.

Once control over the HVAC system has been established (link 3 in Fig. 2), the attacker develops a malware client designed to receive and interpret thermal signals from the environment (sent from the local air conditioner).

This client-side receiver is not difficult to make since many personal computers today include a thermal sensor. The computer's BIOS collect these readings and present them to the operating system. For instance, the Win32_TemperatureProbe class from Window's WMI Provider gives live temperature readings of the host's CPU. Once the malware has control over the host, these sensors can be exploited to sense changes of temperature from the environment (discussed later in more detail). A command interpreted from the thermal signals will cause the malware to perform a pre-coded task based on the received parameters (such as those from Table I).

Once the malware is ready, the attacker performs stage 2 the APT from Fig. 1 by infecting some of the hosts in the target network using one the methods mentioned in Subsection 2.1. Lastly, APT stages 3 and 4 are performed autonomously as the malware spreads across the network gaining access to various assets.

At this point, the attacker has a unidirectional broadcast channel to the infected hosts to command them when needed (links 4 and 5 in Fig. 2). Although the attacker does not know which hosts within the building have been successfully infected, the attacker has control over the building's HVAC system and can perform an effective broadcast to all locations. Since the malware propagated autonomously, the probability of successfully making contact is increased. Moreover, the attacker has the option to design the malware clients to keep a low signature communication network among themselves.

Doing so will help strengthen the foothold on the network by enabling the propagation of newly received commands to hosts that are not affected by the changes in the temperature.

Now the Attacker has all the needed tools to perform various attacks at stage 5 of the APT in Fig. 1 with flexibility in timing. In order to avoid suspicion, the attacker will broadcast of the commands (i.e., the changing of the rooms' temperature) at night when no personnel are present.

III. THE COMMUNICATION CHANNEL

One way of transmitting information is to use the air itself. Essentially, communication is the act of propagating signals across a physical media over some distance. If an air conditioner (the transmitter) emits heat signals from a publically networked source to a thermal sensor of a computer (the receiver), then the attacker has affectively bridged the air gap.

The subject of channel line-encodings (the method of which binary is represented and propagated over a physical medium) has not been researched in depth for thermal channels. The commonly used line-encoding schemes are designed for specific kinds of physical media [30]. Voltage and electro-optical-based line-encodings are not suitable for transmitting binary data thermally across open environments because of the slow turnaround time from hot to cold. For example, the classic return to zero (RZ) encoding technique and all encodings from the same family require a full transition from one level to another to represent a bit. Modulation schemes such as phase-shift keying (PSK), frequency-shift keying (FSK), and quadrature amplitude modulation (QAM) all depend on a carrier wave, making them problematic for the same reason.

In this section we propose a new thermal line-encoding, suitable for such environments, which can be used in the proposed adversarial attack model. Although the data-rate is low (with respect to modern communication channels) it is more than enough for the attacker's needs according to the attack scenario (further details on the data-rate can be found in Section 5).

Lastly, although we outline a thermal channel line-encoding, there are many different variations that can achieve the same goal. This section should serve as a guide for security experts in helping them know what to look for and what is possible with such an attack.

A. Covert Thermal Line-Encoding

The most important characteristic of the proposed channel is its covert capabilities [31]. In other words, the secrecy of the channel is more important than its capacity for transferring bits. Although the attacker from this scenario (Section 2) will only transmit bits during the night, it is still a good idea to maintain a low profile. There are many ways to achieve this; we will offer one method to demonstrate its plausibility.

In our encoding scheme, we assume that it is impossible for the temperature to be the same or to change at the same rate for all areas within the environment. We believe this to be a reasonable assumption since the actual room's temperature will vary depending on the proximity of the thermal sensor (receiver) to the source of the air conditioning unit.

The proposed encoding forms a non-linear time invariant

(NTI) system that is similar to the RZ line coding scheme. Similarly to RZ, the observable signal (temperature) fluctuations are minimal. In other words, in RZ, if there is a series of ones or zeroes—then the signal remains unchanged. As mentioned earlier, this imposes a problem in synchronization since a long stream of the same bits can cause timeslots to slip. This is why the proposed protocol (addressed later) limits the frame length, and includes a short synchronization preamble.

Let the proposed simplex channel be $C$ and its transmitter and receiver be $A$ and $B$ respectively. Let $T$ be the length of the timeslot and equivalently the transmission time of a single bit. Let $H$ and $L$ be the maximum and minimum target transmissions temperatures of the encoding in Celsius, and let $D = |H - L|$ be their difference. Let $O$ be the significant difference in temperature in Celsius which is observable by an unaware human subject over the interval $T$. Although the attack scenario from Section 2 takes place at night and assumes that there are no personnel present, we include the notion of $O$ to help extend the channel to other possible scenarios. Let $\gamma$ be the receiver's thermal sensor's resolution (sensitivity). Finally, let $M(A, B)$ be the impact function (or rate of temperature affect) of the thermal source(s) of $A$ with respect to the location of $B$'s thermal sensor within the environment.

When initiating a transmission, $A$ should select $H$ and $L$ with respect to the current room temperature. Furthermore, the selection should be made such that $D$ will be minimal in respect to $O$ and $\gamma$. From the perspective of the attacker, it is difficult to know any prior information about the $\gamma$ and $M(A, B)$ of the target. Therefore, it is a good idea to allot as much flexibility to $\gamma$ as possible with respect to $O$.

Since it is assumed that there are no prior configuration settings shared between the two parties, the encoding is observation-based. This means that a bit is decoded by comparing it to the previously received bit (or signal level), and that no prior channel configurations are needed (aside from $T$). In tables II and III, the $A$ encoding rules and $B$ decoding rules can be found respectively.

TABLE II
THE TRANSMITTER'S LINE-ENCODING RULES

|  | Target Temperature |
|---|---|
| Bit to transmit is 0 | L °C |
| Bit to transmit is 1 | H °C |

TABLE III
THE RECEIVER'S LINE DECODING RULES

| Previously Received Bit: | 0 | 1 |
|---|---|---|
| Current trend: °C↑ | 1 | 1 |
| Current trend: °C → | 0 | 1 |
| Current trend: °C ↓ | 0 | 0 |

In Table II, it is clear that the transmitter is only concerned with whether it should continue to approach the target temperature ($H$,$L$) or not. In contrast, Table III shows that the receiver takes into account whether the transmitter has reached a target level by considering the non-changing trend °C →.

Note that $H$ and $L$ are the target values and not the maximum values. For instance, although the air conditioning units within the HVAC system are capable of reaching more extreme temperatures, in the scenario where $A$ is an air conditioning unit, $H$ could be 26 °C and $L$ could be 23 °C. Limiting the bounds on the temperatures is important in order to ensure that $O$ is reasonable. Furthermore, in some cases, a very low $L$ can be problematic in that it could take a long time for $B$'s chassis to heat up. This would lead to $B$ missing several symbols or even an entire message.

In order for this line coding scheme to work, it is imperative that some threshold $\mu$ be considered for a trend to be called non-changing. However, it is advantageous to note that $\mu$ can be rather small after noise filtering has been applied to the received signal.

### B. Sensor Noise Mitigation

It is obvious that not all computers have the same $\gamma$. For instance, the precision of the computer's internal temperature readings are dependent on the API available to the software. Usually, these temperature readings are represented as integers (and not as floats). Therefore, the recorded samples from the quantization process have an inaccuracy due to the incurred rounding. Consequently, the detection of low temperature shifts within the range of $O$ is very difficult to attain. This in turn lowers the channel's bit rate since it takes a much longer $T$ for a change in environment temperature to be detected with certainty.

Fortunately, the quantization problem can be remedied by using a moving average filter (MAF) [32]. The MAF is a sliding window which averages the last observed discreet samples into a single time reference. By using an MAF it is possible to achieve an accurate reading of the environmental temperature effectively. This is especially the case considering the relatively high sampling rate, available on most modern devices. The MAF noise mitigation algorithm for some time span is presented in Algorithm 1. The performance of the algorithm is available in Section 5.

---

**Algorithm 1:** Moving Average Filter for Noise Mitigation

---

**Input:** The array of sampled temperatures "$T$",
       the window size parameter to average over "$w$",
**Output:** A smoother representation of $T$ in the array "$S$"
**Begin:**
    1. FOR $i = 1$ TO $Length(T) - w$
    2. $t \leftarrow Average(\ T[i\ to\ (i + w)]\ )$
    3. $S[i] \leftarrow t$

---

On lines 1-3 of Algorithm 1, an array of sampled values are smoothed using a moving average, with a window size of $w$. From Algorithm 1, it is clear that parameter $w$ causes a delay in the channel's output. The optimal (smallest possible) parameter for $w$ is dependent on the channel's bit rate and the sampling frequency. Note, it is preferable in general to have a high sample rate since more information is captures and changes in the signal can be more rapidly detected.

## IV. COVERT THERMAL TRANSMISSION PROTOCOL

Having proposed the line coding, we will now discuss the transmission protocol.

It is necessary to implement a transmission protocol since a covert channel requires many considerations. For instance, in general an attacker will prefer to generate as little noise as possible to avoid detection. This is also true in the case where assumedly nobody is pre-sent. Therefore, the attacker will ensure that the cumulative transmission time will be minimal. Furthermore, in order to avoid frequent changes in the observable room temperature, the attacker will design the frame format accordingly and with care.

Another physical layer protocol rule the attacker will consider is the mitigation of corrupt or falsely sent messages from being accepted. For example, if the local personnel legitimately leave a door open to an office before going home. In this section, we discuss the transmission protocol that complements the channel from Section 3. We will also propose a possible frame format that an attacker may use in order to address the aforementioned challenges.

### A. Frame Format

The proposed protocol is a frame based messaging broadcast. The frame structure can be seen in Fig 3. At the start of the frame, a preamble of "10" is sent. After the preamble, it is assumed that the receiver is actively decoding the temperature trends according to Table III. At this point a 3 bit OP-code is transmitted along with an optional parity bit that covers the Op-code and the payload. Instead of a parity bit, a forward error correction code (FEC) such as hamming code [32] can be used. We found in our experiment that a parity bit was sufficient to form a reliable channel. After the parity bit, a payload of n bits is transmitted.

| 2 bits | 3 bits | 1 bit (optional) | n bits |
|---|---|---|---|
| Preamble "10" | Op-code | Parity bit | Payload |

Fig. 3. The covert thermal transmission protocol's frame format.

### B. Protocol

From the perspective of the transmitter, the preamble has a particular significance. The "10" indicates that the room temperature should be raised (for as long as needed) to get to a generally readable state, and then reduce the temperature over the time T. This is done because the temperature sensors of the common PCs we encountered do not read values below a certain range.

Moreover, the preamble is advantageous in its ability to decouple H and L from all previous transmissions. For instance, let us assume that originally the room temperature starts low and one frame is sent successfully. Afterward, the local personnel legitimately change the temperature level to a higher level. The transmitter and receiver can process the next frame with no ties to the previous room temperature measurements.

Note that in order to mitigate the case of a falsely received preamble, we suggest that the receiver both examines/verifies the parity bit in the frame and follows a state machine to catch any possible "true" preambles received in the middle of a false transmission. The proposed protocol is only an example. Should a higher reliability be necessary, then forward error correction codes (FEC) may be used instead of a parity bit. However, it is worthwhile to mention that FECs, such as hamming code, may not be suitable. This is because the channel may be "sluggish" when changing temperatures near L and FECs, like hamming code, require an equal likelihood between receiving a '1' or '0' in order to work.

As for the OP-code field, we believe that only a few operational commands are required for the purposes of controlling an APT or other similar infiltrations in accordance with the uses listed in Table I. Examples of possible Op-codes are: "Disable Assets of type X", "Delete all evidence", "Search and Destroy file Y", and so on. The frame's payload is self-explanatory. Depending on the Op-code, it could be anything from a file name to an asset type, or even a code obfuscation seed.

The length of the payload (n) depends on the Op-code. For instance, a self-destruct command may be 1 bit while a change in encryption key may be 128 bits or more.

In order to minimize the error-rate, an attacker may have the air conditioner warm up its components before any transmission by heating and then cooling the room. Moreover, in order to limit the number of incorrectly read frames, the attacker may have designed the protocol to transmit and receive frames exactly after a certain set time (e.g., 2:00 a.m.) and not just "anytime at night." In this case the air conditioner may warm up its components exactly prior to that time.

Note that there is a possibility of peer collaboration between infected hosts. Although traffic within the isolated network may raise alarms of an intrusion detection system (IDS) [33], a botnet within an organization can go undetected if the traffic is shaped intelligently [34]. This gives an added advantage to the scenario depicted in Section 2 because there is a high probability that many of the infected hosts have either very "noisy" thermal channels with the local air conditioning unit in their room, or are simply out of range. A peer to peer collaboration network can be used to help pass an attacker's command to the infected host holding the target resource (or near the target asset).

Furthermore, due to the broadcasting nature of the channel, it can be assumed that all receivers should get the same messages sent by the HVAC system. Hence, it is possible to boost the reliability of the message quality if the infected peers share their information. For instance, after the parity bit check, a majority vote on a received frame can correct the corrupted frames.

An example UML state chart of the communication protocol of all the coded into malware from the attack scenario (Section 2) can be found in Fig. 4. This figure demonstrates the usage of all the concepts of the protocol discussed thus far.

## V. EXPERIMENTAL RESULTS

In this section we present experimental results in order to concretely show that a covert thermal channel from an air conditioner to a nearby computer is possible. The results demonstrate that

a) changes in room temperature caused from a conventional air conditioning unit (within an HVAC system) are detectable by a common desktop computer, and

b) the proposed line-encoding and frame format can be used to reliably transmit over an air gap.

The computer used as the receiver in all experiments was an ordinary PC having the specifications in Table IV.

## A. Line-Encoding Performance

There are two aspects which we tested in order to evaluate the line encoding's performance; the noise reduction of the MAF on the quantized signal samples, and the impact thermal

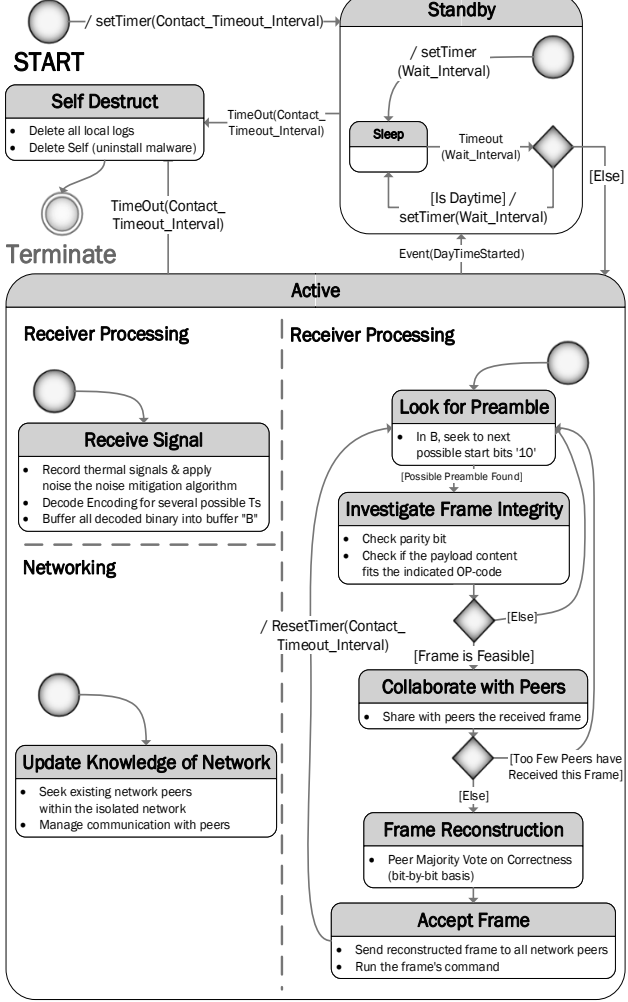

Fig. 4. An example UML state machine for the proposed communication protocol. The "[ ]" brackets symbolize a condition, "/" indicates an action taken, and dashed lines represent parallel states.

emissions (such as those incurred from casual CPU usage) have on thermal channel.

The PC used in the experiment had a sample resolution ($\gamma$) of 1°C. This caused a significant amount of noise, in particular when trying to detect short signal impulses. Fig. 5 shows how the noise mitigation algorithm (MAF) significantly smooths the noise in the transmitted signal impulse received by a PC's CPU thermal sensor. For example, a signal impulse (heating on/off) is illustrated as a thick grey line in Fig. 5, where a nearby air conditioner (3 meters away) heats the room for 90 seconds and is then turned off. A step response (depicted as a blue line in the same figure) is how the impulse signal sent from the transmitter (air conditioner) is received by the receiver (the PC's thermal sensor). The step response is useful since it can be used to measure various properties of the channel, such as the channel's noise, propagation delay and bit-rate.

We note that in our experiments, an open chassis causes a slower rate of ascent and a faster rate of descent in contrast to having a closed chassis. This result implies that closed chassis computers can receive bits at a faster rate than open chassis computers (e.g., open server racks).

Fig. 6 shows the impact of casual usage of the receiving computer on the malware's capability to receive environmental thermal signals. Note that the thermal sensor located on the main electronic board of the computer (motherboard) is insignificantly affected by running processes. This means that thermal signals transmitted over the environment will have little interference from the heat emitted from the local computer's components. We use this thermal sensor to evaluate the performance of the communication protocol.

## B. Protocol Performance

To test the protocol, we generated a possible frame an attacker might broadcast to all receiving computers un-der the same HVAC system. This frame consists of a "Change Encryption Key for Internal Communications" OP-code and a payload with the new 128 bit key (other sized keys are possible). In other words, assuming that there is internal cooperation between infected hosts, all infected hosts which receive this command will update their channel encryption keys accordingly.

We found that the thermal sensor located on the motherboard responded faster to changes in the environment' temperature than the CPU's thermal sensor. This is understandable due to the cooling unit (fan and heat sink) attached to the CPU. For this reason, we use the motherboard's sensor for testing the protocol.

First analyzed the step response of the air conditioner going from H to L on the motherboard's thermal sensor where H was 26 °C and L was 23 °C. From these results (Table V), we decided to use a T of 1.5 minutes for the protocol test and set the other parameters accordingly (Table VI). With this setting a frame of 134 bits could be transmitted in about three a half hours (overnight) with a bit rate of 40 bits per hour (bph). Afterwards we transmitted the 134 bit frame while the receiver sampled the sensor (Fig. 7) and then demodulated the bits (Fig. 8).

It should be noted that the bit rate of the channel will be better or worse depending on the scenario (such as the receiver's distance from the transmitter). However, we only tested a receiver sample rate of 3.3 Hz. With a much higher sample rate and the same quantization $\gamma$, it may be possible to achieve an even higher bit rate after applying the MAF.

Regardless, it is conceivable that 40 bph is more than enough for the attacker to trigger a task from Table I during the night hours (refer to the scenario in Section 2).

Overall the proposed protocol and line coding were successful in our experiments, and have demonstrated the plausibility of such a channel.

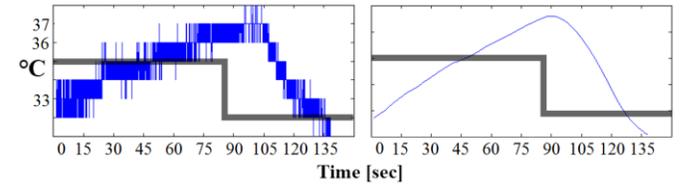

Fig. 5. The air conditioning unit's impact (step response) on a closed chassis PC, 3 meters away. Left: the integer samples from the CPU's thermal sensor obtained from the API. Right: the same samples after applying the noise mitigation algorithm.

TABLE IV
PARAMETERS TAKEN FOR TESTING THE PROTOCOL

| Parameter | Value |
|---|---|
| *Case Type* | Mini ATX |
| *Power Supply* | 280W LiteON PS-4281-02, 240W input |
| *Internal Fans* | 1 Case Fan, *Asia Vital* CPU Fan |
| *Motherboard* | Lenovo Mahobay, 0C48431 Pro |
| *CPU* | Intel i5 3470 @ 3.20Ghz |
| *RAM* | Transcend 2x4GB DDR3 800MHz |
| *Hard-drive* | Hitachi HDP725050GLA360 ATA Device 500GB |
| *Operating System* | Windows 8 Enterprise Build 9200 |

TABLE V
THE AIR CONDITIONING UNIT'S IMPACT ON
THE MOTHERBOARD'S SENSOR AFTER NOISE MITIGATION

| | |
|---|---|
| *Lowest Recorded Temperature* [℃] | 22 |
| *Highest Recorded Temperature* [℃] | 30 |
| *Maximum Temperature Difference* [℃] | 8 |
| *Distance from Transmitter* [m] | 3 |
| *Dimensions of the Office Room* [m³] | 4x4x5 |
| *Linear Rate of Ascent* [℃/min] | 1.23 |
| *Linear Rate of Descent* [℃/min] | -1.24 |

TABLE VI
PARAMETERS TAKEN FOR TESTING THE PROTOCOL

| Parameter | Value |
|---|---|
| $C$ | Small office room |
| $t$ | Centralized AC system |
| $r$ | Closed chassis desktop computer |
| $H$ | 26 ℃ |
| $L$ | 23 ℃ |
| $D$ | 3 ℃ |
| $\gamma$ | 1 ℃ |
| $\mu$ | 0.01 ℃ |
| **MAF window** | *1 minute of samples* |
| **Sample Rate** | *3.3 Hz* |

## VI. FORENSICS & COUNTERMEASURES

There are various ways that the adversarial attack model presented in this paper can be detected and pre-vented. These ways can be seen in the attack steps presented in Fig. 2, and are enumerated in this section.

### A. HVAC Internet Connectivity

The most direct way to thwart the attack is to increase the security around the HVAC system. In particular maintenance personnel should be trained to be security aware, by being careful about their passwords, and not to leave their infrastructure systems unencrypted. Moreover, they should know what to keep an eye out for in order to report suspicious activity.

Another option is to place a strong firewall in front of the web portal to the system. However, considering the reports of vulnerabilities in HVAC systems such as Ni-agara [19], a firewall may be of little help, especially in the case of an APT. Therefore, the best way to secure an air gapped network is to have all overlapping networks air gapped as well, and disconnect the HVAC system completely from the internet.

### B. Monitoring the Thermal Environment

In the case where disconnecting the HVAC from the internet is not an option, a less direct way to detect the channel is to monitor the thermal environment. This can be done by placing thermal sensors in various rooms, and by recording the temperature fluctuations. If the channel is used at night, then it will be exposed considering there are no personnel around (assumedly).

This strategy is even more affective in rooms where the temperature generally stays at the same level. For ex-ample, a server room's temperature is closely regulated 24/7 to stay at a low level. Consistent and correlated rises and drops on temperature at any point in the day may indicate the presence of a covert thermal channel. This requires further research as to the protocol or ma-chine learning algorithm that can be used to detect these anomalies, while providing a low false alarm rate.

### C. Malware Signatures

In order for an infected host to detect and receive information over a thermal channel, it must consistently sample the available thermal sensors. After which, it can be expected that some typical noise mitigation processes will be performed on the received samples.

This type of behavior can be detected by performing static and dynamic analysis on the code. Such a signature would involve a high frequency access to the OS's thermal sensor APIs.

If code is found having the capability of the de-scribed behavior, then the presence of the covert thermal channel is exposed. Therefore, it is worthwhile tracking these API calls, especially since they do not require privileged permissions making (it is feasible for a non-administrator user to install the malware on a targeted host).

## VII. CONCLUSION & FUTURE RESEARCH

The popular separation of networks by use of an air gap is susceptible to attacks from parallel insecure networks. In this paper we have shown a new adversarial attack model, where the attacker broadcasts commands to infected hosts within an isolated network. We have also provided a plausible scenario in which an attacker might use this model. Moreover, through discussion and experimentation, we have demonstrated the feasibility of signaling a computer from an air conditioner using a new thermal line-encoding and complimentary communication protocol. We encourage network engineers and security experts alike to take this attack into consideration when designing an air gapped network.

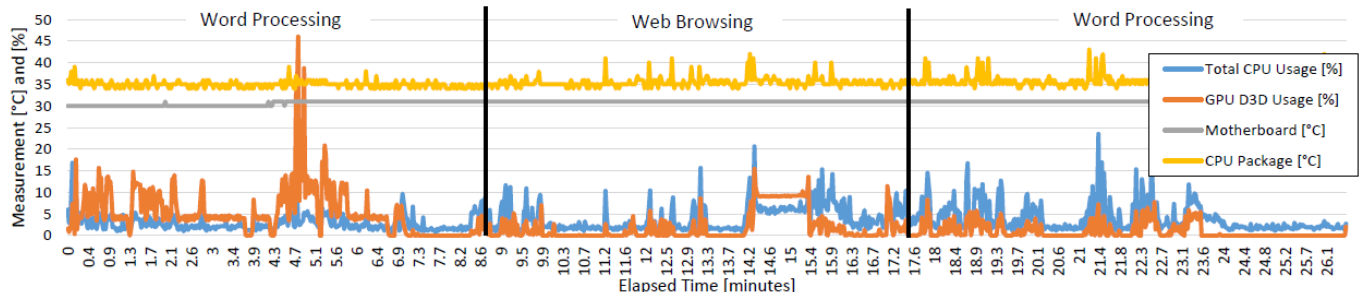

Fig. 6. The impact of the receiving computer's casual usage on the thermal sensors.

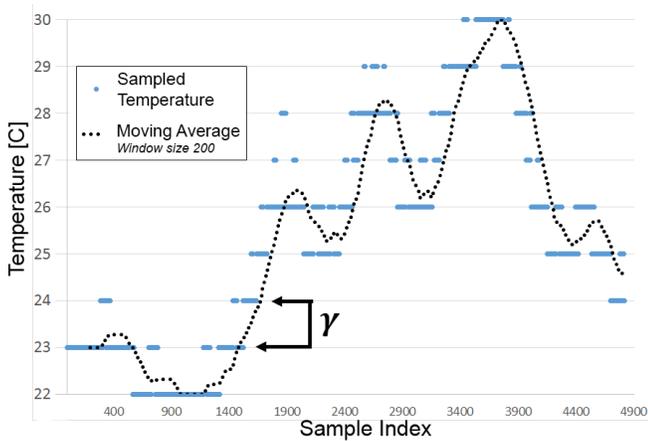

Fig. 7. The start of a receiver's reception of a 138 bit frame. First the sensor is sampled, and then a noise mitigation filter (MAF) is applied. Note that although the transmitter (air conditioner) only targets the temperatures 23 °C and 26 °C, the temperature inside the chassis of the receiver is higher.

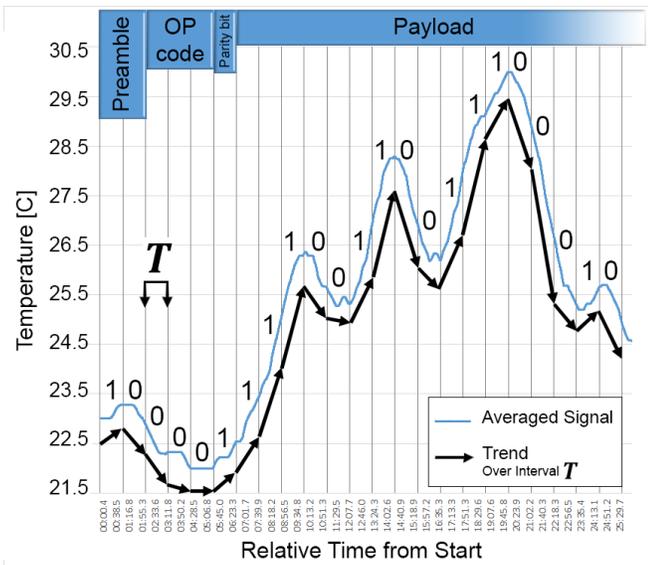

Fig. 8. After applying the noise mitigation, the signal is demodulated into a frame by observing the trends over each interval. The intervals are defined by the shape of the preamble. Here the OP-code is "000" and the payload is a 128 bit encryption key (only a partial frame is shown).

Although we have shown how a 40 bit per hour rate channel is achievable, we believe that even faster bit rates are possible. For instance, we performed our experiment by using the motherboard's heat sensors. There are many more heat sensors available (such as the CPU, GPU and Hard-drive) which can be used to create a single input multiple output (SIMO) channel that maximizes the channel's capacity. Moreover, the device's exhaust fan speed can be used as well (after applying a MAF) since it is correlated to the environment's temperature.

Furthermore, we intend to reconstruct the experiment over a server farm. Such an attack can be used to cause data leakage or another type of sophisticated attack. For instance, it is conceivable to thermally transmit messages from one server to another in the same server rack as mentioned in [25]. We intend to develop a thermodynamic computer model that will allow us to run more simulations, efficiently.

Moreover, an attacker may wish to have an open channel during the daily working hours. We intend to expand the proposed protocol to take into account the presence of people in the room. This is particularly challenging since presence not only has an influence on the PCs' temperature (quality of signal), but it also makes the covert channel more susceptible to detection (considering the parameter O). Not only would it be interesting to see if an attacker could establish a covert channel during the day, but determining its plausibility may also help detect such attacks in the future.